\documentclass{elsart}
\usepackage{amsmath}

\begin{document}
\begin{frontmatter}
\title{Model-free derivations of the Tsallis factor: 
                  constant heat capacity derivation}
\author{Wada Tatsuaki\corauthref{cor1}}
\ead{wada@ee.ibaraki.ac.jp}
\address{Department of Electrical and Electronic Engineering, 
Ibaraki University, Hitachi,~Ibaraki, 316-8511, Japan}
\corauth[cor1]{Corresponding author.}

\begin{abstract}
The constant temperature derivation, which is a model-free
derivation of the Boltzmann factor, is generalized in order to develop
a new simple model-free derivation of a power-law Tsallis factor based
on an environment with constant heat capacity.
It is shown that the integral constant $T_0$ appeared in the new derivation
is identified with the generalized temperature $T_q$ in Tsallis 
thermostatistics.  
A constant heat capacity environment is proposed as a one-real-parameter
extension of the Boltzmann reservoir, which is a model
constant temperature environment developed by J.~J.~Prentis {\it et
al.} [Am. J. Phys. {\bf 67} (1999) 508] in order to naturally obtain
the Boltzmann factor. It is also shown that the Boltzmann entropy
of such a constant heat capacity environment is consistent with Clausius'
entropy.
\end{abstract}

\end{frontmatter}

\section{Introduction}
\label{intro}
One of the most important factors in science is the Boltzmann factor,
which governs the thermal behavior of any system in nature at constant
temperature $T$.  It is well known that when a system is in
equilibrium with its environment, a probability that the system is in
an accessible microstates of energy $E_s$ is proportional to the
celebrated Boltzmann factor, $\exp(-\beta E_s)$, in the limit that the
environment becomes a true heat reservoir. In this {\it reservoir
limit}, $\beta$ is identical to the environment's inverse-temperature
$\beta = 1/(k T)$, which is defined by
\begin{equation}
  \beta \equiv \frac{\partial \ln \Omega(U)}{\partial U},
  \label{Def-beta}
\end{equation}
where $\Omega(U)$ is the number of accessible microstates of the
environment at the energy $U$, and $k$ is Boltzmann constant.  A true
heat reservoir can be defined by an environment whose temperature is
exactly constant irrespective of its energy gains or losses.  Recently
Prentis, Andrus, and Stasevich \cite{pren99} have closely studied the
precise conditions that generate the Boltzmann factor emerge naturally
by examining the exact physical statistics of a system in thermal
equilibrium with different environments.  They proposed the Boltzmann
reservoir (BR), which is a model constant-temperature environment, in
order to develop new and improved ways of obtaining the Boltzmann
factor.  For any system in thermal contact with a BR, the equilibrium
distribution is identical to an exponential Boltzmann
distribution. The Boltzmann factor naturally emerges without any
assumptions about constant temperature, and without resort to any kind
of {\it reservoir limit}.  The BR is thus a true heat reservoir, and
its temperature is exactly constant independent of the amount of
energy it gains or losses.  The interesting and nontrivial properties,
such as non-concavity of entropy, a non-invertible Legendre
transformation, inequivalence of the canonical and microcanonical
ensembles, etc., of BR have further studied by H. S. Leff
\cite{leff00}.

On the other hand, there has been growing interest in the nonextensive
generalizations of the conventional Boltzmann-Gibbs (BG) statistical
mechanics.  One of them is the nonextensive thermostatistics
\cite{tsal88,tsal99,tsal02} based on Tsallis' entropy
\begin{equation}
  S_q = k \frac{1 - \sum_i p_i^q}{q - 1},
\hspace{1cm}(\sum_i p_i=1; \hspace{3mm} q\in {\mathcal R})
\label{Sq}
\end{equation}
which is a nonextensive extension of the conventional BG entropy by
one-real-parameter of $q$.  In the limit of $q \to 1$, Tsallis'
entropy Eq. (\ref{Sq}) reduces to BG entropy $S_1 = -\sum_i p_i \ln
p_i$.  The maximization of $S_q$ with respect to $p_i$ under the
constraints imposed by the normalization and the energy $q$-average
$U_q = \sum_i E_i p_i^q / \sum_i p_i^q$ yields a power-law probability
distribution
\begin{eqnarray}
  p_i &\propto& \left\{1-(1-q) \frac{\beta_q (E_i-U_q)}{\sum_j p_j^q} 
             \right\}^{\frac{1}{1-q}}
      = \exp_q[-\frac{\beta_q (E_i-U_q)}{\sum_j p_j^q}],
\end{eqnarray}
where $\beta_q$ is the Lagrange multiplier for the constraint 
associated with the energy $q$-average $U_q$, and
\begin{equation} 
  \exp_q(x) \equiv \{1+(1-q)x \}^{\frac{1}{1-q}},
  \label{q-exp}
\end{equation}
is the $q$-exponential function, which reduces to $\exp(x)$ in the
limit of $q \to 1$.  We name the factor $\exp_q(-{\tilde \beta} E_s)$
Tsallis factor, where ${\tilde \beta}$ is a quantity related to the
temperature of the environment in thermal equilibrium.  The Tsallis
factor is a generalization of the Boltzmann factor by the
real-parameter of $q$. It can treat both power-law ($q \ne 1$) and
exponential-law ($q=1$) distributions on an equal footing.
M.~P.~Almeida \cite{alme01} has derived Tsallis
distribution if we assume that (the inverse of) the heat capacity 
of the environment is related by
\begin{equation} 
 \frac{d}{dE}(\frac{1}{\beta}) = q-1.
\end{equation}
The remarkable point is that Tsallis parameter $q$ is given
a physical interpretation in terms of heat capacity of the environment.
For a finite heat capacity ($q \ne 1$) we obtain power-law distribution,
while for an infinite heat capacity ($q=1$) we recover canonical 
BG distribution.

In this paper, inspired by the idea of J.~J.~Prentis {\it et al.} \cite{pren99}
 and that of M.~P.~Almeida \cite{alme01},
the model-free derivations of the Boltzmann factor are
generalized in order to obtain the Tsallis factor, which is a
one-real-parameter extension of the Boltzmann factor.  
The derivations of the Boltzmann factor are based on a constant-temperature
environment (heat reservoir), whereas the
generalized derivations of the Tsallis factor are based on a
constant heat capacity environment.
It is also shown that Almeida's method \cite{alme01} is considered as 
a generalization
of small-$E_s$ derivation, which is one of the model-free derivations
of the Boltzmann factor. A new derivation is developed by generalizing
the other model-free derivation (constant-$T$ derivation) of Boltzmann
factor. 
  
The rest of the paper is organized as
follows: in the next section we briefly review the model-free
derivations of the Boltzmann factor (the constant-$T$ and small-$E_s$
derivations). It is emphasized that the Boltzmann factor is not the
unique factor which is obtained in the small-$E_s$ derivations;
in section III it is shown that the Tsallis factor can be obtained 
by extending the two model-free derivations of the Boltzmann factor. 
Then the constant heat capacity derivation is proposed as a one-real-parameter
extension of the constant-$T$ derivation; in section IV, after the brief 
explanation of BR, we propose a constant heat capacity environment,
{\it Tsallis reservoir} (TR), as a generalization of BR. It is also shown
that the Boltzmann entropy of TR is consistent with the Clausius' definition 
of thermodynamic entropy. The integral constant $T_0$ in the constant heat
capacity derivation is shown to be identical with the generalized temperature,
which is the thermal conjugate quantity of Tsallis' entropy. 
Final section is devoted to concluding remarks.

\section{Model-free derivations of the Boltzmann factor}
Let us consider a system plus its environment as an isolated system.
The total energy $E_t=E_s + U$ is conserved and as a consequence the
environment has an energy $U=E_t-E_s$ when the system is in a
microstate of energy $E_s$. According to the equiprobability postulate
in statistical mechanics, each accessible microstates of the
system-plus-environment is equally probable. The probability $p_s$ of
the system in a state of energy $E_s$ is thus proportional to the
number of accessible microstates of the environment, i.e.,
$p_s \propto \Omega(E_t-E_s)$.  It is worth noting that the number of
energy-shifted microstates $\Omega(E_t-E_s)$ is the central object
which uniquely determines the probability $p_s$ of each microstate of
the system.

The standard Boltzmann factor can be derived irrespective of the
microscopic nature of environments. Such model-independent derivations
can be categorized into two main classes \cite{pren99}: the
constant-$T$ derivations; and small-$E_s$ derivations.  The
constant-$T$ derivations assume that the temperature $T$ of the
environment is exactly constant, hence the environment is a true heat
reservoir. Under the constant-$T$ condition, we readily obtain the
Boltzmann factor as $\Omega(E_t-E_s) = \Omega(E_t)\cdot \exp(-\beta
E_s)$ by integrating Eq. \eqref{Def-beta} from $U=E_t$ to
$U=E_t-E_s$. However there is no real physical environment which has an
exactly constant temperature.  Thus here comes the small-$E_s$
derivation, in which it is assumed that an energy $E_s$ of the system
is sufficiently small compared to the total energy $E_t$, i.e., $E_s
\ll E_t$.  The first step of the small-$E_s$ derivation is rewriting
the number of energy-shifted microstates by utilizing the pair of
exponential and logarithmic functions,
\begin{equation}
       \Omega(E_t-E_s) = \exp \left[ \; \ln \Omega(E_t-E_s) \; \right].
  \label{exp-log-rep}
\end{equation}
Then the term $\ln \Omega(E_t-E_s)$ is expanded in terms of small
$E_s$ around $E_t$,
\begin{equation}
   \ln \Omega(E_t-E_s) = \ln \Omega(E_t) 
             - \frac{\partial \ln\Omega(E_t)}{\partial E_t} \cdot E_s + \cdots.
\end{equation}
By keeping only first-order term in $E_s$ we obtain the Boltzmann factor
\begin{equation}
   \Omega(E_t-E_s) = \Omega(E_t) \cdot \exp(-\beta E_s).
\end{equation}
This is the traditional approach to obtain the Boltzmann factor.  At
first sight it seems that the Boltzmann factor uniquely emerge from
the small-$E_s$ derivation.  However this is not true! Indeed, S.~Abe
and A.~K.~Ragagopal \cite{abe-raja01,abe-raja01-euro} have shown 
the non-uniqueness of Gibbs' ensemble theory.  
They also pointed out that the choice of the
pair functions other than exponential and logarithmic functions for
rewriting the number of energy-shifted microstates
Eq. \eqref{exp-log-rep} is also possible.
According to them \cite{abe-raja01,abe-raja01-euro}, we review 
that another choice of the pair
functions leads to a different factor within the same framework of the
small-$E_s$ derivations.  For the sake of notational simplicity, let
me introduce the ${\mathcal Q}$-exponential function, 
\begin{equation} 
  \exp_{\mathcal Q}(x) \equiv (1+{\mathcal Q} x)^{1/{\mathcal Q}},
  \label{Q-exp}
\end{equation}
and its inverse function, the ${\mathcal Q}$-logarithmic function,
\begin{equation} 
\ln_{\mathcal Q}(x) \equiv \frac{x^{\mathcal Q} -1}{\mathcal Q}.
\end{equation}
In the limit of ${\mathcal Q} \to 0$, these functions reduce
to ordinal exponential and logarithmic functions respectively.
Note that we can define some variants of the $q$-exponential function of
Eq. \eqref{q-exp} and their inverse functions by specifying ${\mathcal Q}$
as a function of Tsallis' entropic index $q$, e.g., ${\mathcal Q}=q-1, 
{\mathcal Q}=1-q$, or ${\mathcal Q}=q-q^{-1}$, \dots.
The following arguments are valid for any function 
${\mathcal Q}(q)$ of $q$, satisfying ${\mathcal Q}(1)=0$.

By utilizing the pair of the ${\mathcal Q}$-exponential and ${\mathcal
Q}$-logarithmic functions, the number of energy-shifted microstates
can be written as
\begin{equation}
  \Omega(E_t-E_s) = \exp_{\mathcal Q}[ \; \ln_{\mathcal Q} \Omega(E_t-E_s)\; ],
\end{equation}
instead of Eq. \eqref{exp-log-rep}.  By expanding $\ln_{\mathcal Q}
\Omega(E_t-E_s)$, and keeping up to the first order in $E_s$ as
before, we have
\begin{eqnarray}
    \ln_{\mathcal Q} \Omega(E_t-E_s) 
       &=& \ln_{\mathcal Q} \Omega(E_t) 
         - \frac{\partial \ln_{\mathcal Q}\Omega(E_t)}{\partial E_t} \cdot E_s 
         + \cdots \nonumber \\
       &=& \ln_{\mathcal Q} \Omega(E_t) - \beta_{\mathcal Q} E_s + \cdots,
\end{eqnarray}
where we introduce the ${\mathcal Q}$-generalized inverse temperature
\begin{eqnarray} 
  \beta_{\mathcal Q} &\equiv& \frac{\partial \ln_{\mathcal Q} \Omega(E_t)}
        {\partial E_t} 
         = \Omega(E_t)^{\mathcal Q} 
           \cdot \frac{\partial \ln \Omega(E_t)}{\partial E_t}
         = [1+{\mathcal Q} \ln_{\mathcal Q} \Omega(E_t)] \cdot \beta. 
\end{eqnarray}
By utilizing the useful identity of
\begin{equation} 
 \exp_{\mathcal Q}(x+y) = \exp_{\mathcal Q}(x) \cdot 
                  \exp_{\mathcal Q}(\frac{y}{1+{\mathcal Q}x}),
  \label{identity}
\end{equation}
we finally obtain the Tsallis factor,
\begin{equation}
     \Omega(E_t-E_s) = \Omega(E_t) \cdot \exp_{\mathcal Q} (-\beta E_s).
\end{equation}
In this way not only the Boltzmann factor but also the Tsallis factor can
be obtained in the small-$E_s$ derivations.
What we learn here is that the equilibrium distribution obtained by 
the small-$E_s$ derivations is not uniquely determined! 
In addition, the parameter ${\mathcal Q}$ is not determined at all.
In fact, there is no difference until the first-order in $E_s$ between
the both factors:
\begin{eqnarray}
  \exp_{\mathcal Q}(-\beta E_s) &=&
          1 - \beta E_s + \frac{(1-{\mathcal Q})}{2}(\beta E_s)^2 + \cdots, \\
  \exp(-\beta E_s) &=&
          1 - \beta E_s + \frac{1}{2}(\beta E_s)^2 + \cdots.
\end{eqnarray}
The higher-order terms in $E_s$ should be therefore taken into account 
at least in order to distinguish the Tsallis factor from the Boltzmann factor.

\section{Model-free derivations of the Tsallis factor}
In this section the both constant-$T$ and small-$E_s$ derivations 
are generalized
in order to derive Tsallis factor $\exp_{\mathcal Q}(-\beta E_s)$.
M.~P.~Almeida \cite{alme01} has shown that a power-law Tsallis
distribution can be obtained if the heat capacity of the environment
is assumed to be exactly constant. His method of obtaining the Tsallis
factor can be considered as a generalization of the small-$E_s$
derivation, but $E_s$ is not necessarily small. According to him,
let us suppose that the heat capacity of the environment is exactly
constant irrespective of its energy gains or losses,
\begin{equation}
  C_{\rm env} \equiv \frac{d U}{d T} = \frac{k}{\mathcal Q}.
  \label{HC}
\end{equation}
Note that the physical interpretation of the real parameter ${\mathcal Q}$ 
is very clear! It determines the heat capacity of the environment.
The condition of Eq. \eqref{HC} can be identical to
\begin{equation}
     \frac{d}{d U}(\frac{1}{\beta}) = {\mathcal Q}.
    \label{const-HC}
\end{equation}

By the way the ${\mathcal Q}$-exponential function Eq. \eqref{Q-exp} can be 
expanded \cite{yama02} as
\begin{equation}
     \exp_{\mathcal Q}(x) = \sum_{n=0}^{\infty} \frac{{\mathcal Q}_n}{n!} x^n,
\end{equation}
where
\begin{eqnarray}
     {\mathcal Q}_n &=& (1-{\mathcal Q})(1-2{\mathcal Q}) \cdots
           (1-(n-1){\mathcal Q}) \quad \textrm{for} \; n \ge 2,
     \nonumber \\
     {\mathcal Q}_0 &=& {\mathcal Q}_1 = 1.
\end{eqnarray}
From the condition Eq. \eqref{const-HC} and the definition Eq. \eqref{Def-beta}
of $\beta$, we can show the following relation
\begin{equation}
     \frac{1}{\Omega} (\frac{\partial^n \Omega}{\partial U^n}) 
       = {\mathcal Q}_n \beta^n.
     \label{Omega-diff}
\end{equation}
Expanding $\Omega(E_t-E_s)$ in terms of $E_s$ around $E_t$ and 
using Eq. \eqref{Omega-diff}, the summation
over the all order terms can be readily performed as
\begin{eqnarray}
  \Omega(E_t-E_s) &=& \sum_{n=0}^{\infty} \frac{1}{n!} 
                     \frac{\partial^n \Omega(E_t)}{\partial E_t^n} (-E_s)^n
  = \Omega(E_t) \cdot \sum_{n=0}^{\infty} 
      \frac{{\mathcal Q}_n}{n!} (-\beta E_s)^n  \nonumber \\
  &=& \Omega(E_t) \cdot \exp_{\mathcal Q}(-\beta E_s).
\end{eqnarray}
We therefore obtain the Tsallis factor assuming the constant heat capacity
environment but without resort to the small-$E_s$ condition ($E_s \ll E_t$).

Now let us turn on a new model-free derivation of the Tsallis factor. 
As a generalization of the constant-$T$ derivation, we name it 
the {\it constant heat capacity derivation}.
The starting point of the derivation is again Eq. \eqref{const-HC}, i.e.,
the heat capacity of the environment is assumed to be exactly constant.
Integrating Eq. \eqref{const-HC} we obtain
\begin{equation}
  \frac{1}{\beta} = \frac{1}{\beta_0} + {\mathcal Q} U,
  \label{beta}
\end{equation}
where $\beta_0$ is an integral constant.
Substituting Eq. \eqref{beta} into Eq. \eqref{Def-beta}, we have
the following differential equation,
\begin{equation}
  d \ln \Omega(U) = \frac{\beta_0 dU}{1+{\mathcal Q} \beta_0 U}.
\end{equation}
The solution can be written by
\begin{equation}
  \Omega(U) = \Omega_0 \cdot \left\{ 
                 1+{\mathcal Q} \beta_0 U \right\}^{\frac{1}{\mathcal Q}} 
            = \Omega_0 \cdot \exp_{\mathcal Q}(\beta_0 U),
\end{equation}
where $\Omega_0$ is a constant. Note that the inverse temperature of 
the constant heat capacity environment depends on the internal energy $U$,
\begin{equation}
  \beta(U) \equiv \frac{\partial \ln \Omega(U)}{\partial U}
     =   \frac{\beta_0}{1+{\mathcal Q}\beta_0 U},
     \label{beta-env}
\end{equation}
or that the temperature of the environment $T(U) \equiv 1/\{k \beta(U)\}$ 
is energy dependent 
\begin{equation}
  T(U) =  T_0 + \frac{\mathcal Q}{k} U,
  \label{Tenv}
\end{equation}
where $T_0 \equiv 1/(k \beta_0)$.
The number of energy-shifted microstates is written by 
\begin{equation}
  \Omega(E_t-E_s) 
       = \Omega_0 \cdot \exp_{\mathcal Q}[\beta_0 E_t] \cdot
         \exp_{\mathcal Q}[-\beta(E_t) \cdot E_s].
\end{equation}
We thus obtain the Tsallis factor again. The constant heat capacity
derivation is one of the simplest ways to obtain Tsallis factor.
In the limit of ${\mathcal
Q} \to 0$, the constant heat capacity derivation reduces to the
constant-$T$ derivation, since the heat capacity of the environment
becomes infinite and its temperature Eq. \eqref{Tenv} becomes exactly
constant.
It is important that $T_0$ is different from the temperature of the
environment unless ${\mathcal Q}=0$. In the next section 
$T_0$ is identified as the generalized temperature $T_{\mathcal Q}$,
which is the thermal conjugate quantity of Tsallis' entropy.  
The existence of the parameter $T_0$ clearly distinguishes the constant
heat capacity derivation from Almeida's method, which is a generalization
of the small-$E_s$ derivation.
It may shed some light to further study the origin of power-law 
Tsallis distributions.

\section{The Tsallis reservoir}
Let us briefly review the Boltzmann reservoir (BR)
\cite{pren99,leff00}, which is a {\it hypothetical} model environment
whose temperature is exactly constant. The BR enables us to readily
obtain the Boltzmann factor and the canonical ensemble from the
microcanonical formalism.  The BR is originally described in terms of
its energy spectrum,
\begin{equation}
  U(n) = n \epsilon, \quad n=0,1,2,\dots,
\end{equation}
with degeneracy,
\begin{equation}
  \Omega^{\rm BR}(n) = b^n = b^{U/\epsilon},
  \label{original-BR}
\end{equation}
\noindent
where the parameter $\epsilon >0$ is the separation energy between
adjacent degenerate energy levels, and $b>1$ is a dimensionless constant.
Eq. \eqref{original-BR} can be rewritten into the following form,
\begin{equation}
  \Omega^{\rm BR}(U) = \exp(\beta^{\rm BR} U),
  \label{BR}
\end{equation}
where the inverse temperature $\beta^{\rm BR}$ of BR is given by
\begin{equation}
  \beta^{\rm BR} \equiv \frac{d \ln \Omega^{\rm BR}(U)}{dU} 
         = \frac{\ln b}{\epsilon}.
\end{equation}
In order to maintain a strictly constant temperature, the entropy of a BR
must be linear in the internal energy $U$,
\begin{equation}
  S^{\rm BR} \equiv k \ln \Omega^{\rm BR}(U) = \frac{U}{T^{\rm BR}},
  \label{S-BR}
\end{equation}
and its zero-work heat capacity must be infinite \cite{leff00}.

From the number of the energy-shifted accessible microstates 
$\Omega^{\rm BR}(E_t-E_s)$,
the Boltzmann factor emerges naturally without resort to any 
{\it reservoir limit}:
\begin{eqnarray}
  \Omega^{\rm BR}(E_t-E_s) &=& \exp[\beta^{\rm BR} (E_t-E_s)] 
       = \Omega^{\rm BR}(E_t) \cdot \exp(-\beta^{\rm BR} E_s).
\end{eqnarray}

Having described BR, we now propose {\it Tsallis reservoir} (TR),
which is an extension of BR by the one-real-parameter of ${\mathcal
Q}$.  TR is defined by an environment whose number of accessible
microstates obeys the ${\mathcal Q}$-exponential of its internal
energy as:
\begin{equation}
  \Omega^{\rm TR}(U) \equiv \exp_{\mathcal Q}(\beta_0 U)
         \; \propto U^{\frac{1}{\mathcal Q}} \; \textrm{ for large } U,
  \label{TR}
\end{equation}
where ${\mathcal Q}$ is a real parameter which determines the heat capacity
of TR,
\begin{equation}
  C^{\rm TR} \equiv \frac{d U}{d T^{\rm TR}} = \frac{k}{\mathcal Q},
  \label{HC-TR}
\end{equation}
and the temperature of TR is
\begin{equation}
  T^{\rm TR}(U) \equiv \left( \frac{d \ln \Omega^{\rm TR}(U)}{d U} \right)^{-1}
  = T_0 + \frac{\mathcal Q}{k} U.
  \label{T-TR}
\end{equation}
Note that $C^{\rm TR}$ is exactly constant irrespective of its energy gains
or losses. It is worth noting that the index $\beta_0$ is different
from the inverse temperature of the TR unless ${\mathcal Q}=0$. In the
limit of ${\mathcal Q} \to 0$, TR reduces to BR.

For any system in thermal contact with a TR, the thermal equilibrium
distribution is identical to a power-law Tsallis distribution, 
\begin{eqnarray}
  \Omega^{\rm TR}(E_t-E_s) &=& \exp_{\mathcal Q}[\beta_0 (E_t-E_s)]
          = \Omega^{\rm TR}(E_t) \cdot 
                    \exp_{\mathcal Q}[-\beta(E_t) E_s],
\end{eqnarray}
where the identity of Eq. \eqref{identity} and the definition of 
Eq. \eqref{beta-env} are used.

Now let us focus on the relation between the conventional Boltzmann 
entropy of TR and the Clausius' definition of the thermodynamic entropy.
Since TR is an extension of BR and since TR's temperature $T^{\rm TR}$ 
depends on internal energy $U$, it is important to check whether
the both entropies of TR are consistent each other.  
We readily see the derivative of the Boltzmann entropy of TR is equivalent
to the Clausius' entropy as
\begin{equation}
  dS^{\rm TR}(U) \equiv k \; d( \ln \Omega^{\rm TR}(U)) 
             = \frac{dU}{T^{\rm TR}(U)}.
\end{equation}
The important fact distinct from the case of BR is that the temperature
in the Clausius' entropy is linear-$U$-dependent.
Conversely if we assume that the constant heat capacity, or equivalently 
the linear-$U$-dependent temperature $T^{\rm TR}(U)$, and that 
the Clausius' entropy, 
then we obtain the Boltzmann entropy of TR
\begin{equation}
  S^{\rm TR} = \int_0^U \frac{dU}{T_0 + \frac{\mathcal Q}{k} U} 
        = k \ln [\; \exp_{\mathcal Q}(\beta_0 U) \; ].
\end{equation}

On the other hand, if we adopt the ${\mathcal Q}$-generalized Boltzmann 
entropy $S^{\mathcal Q} \equiv k \ln_{\mathcal Q} \Omega^{\rm TR}(U)$ for TR,
we obtain the following relation
\begin{equation}
  S_{\mathcal Q}^{\rm TR} = \frac{U}{T_0},
  \label{Sq-TR}
\end{equation}
which is comparable with the relation of Eq. \eqref{S-BR} for BR. 
We therefore find
that $T_0$ is identical to the generalized temperature $T_{\mathcal Q}
\equiv (\partial S_{\mathcal Q} / \partial U)^{-1}$ and that
the temperature Eq. \eqref{T-TR} of TR is written by
\begin{equation}
  T^{\rm TR} = T_{\mathcal Q} 
     \left( 1+ {\mathcal Q} \frac{S_{\mathcal Q}^{\rm TR}}{k} \right),
\end{equation}
which is equivalent to the relation \cite{abe-mart01} between the physical 
temperature and the $q$-generalized temperature in nonextensive thermodynamics.
Note also that the derivative of Eq. \eqref{Sq-TR} is consistent with
the modified Clausius' definition of thermodynamic entropy  \cite{abe-mart01},
\begin{equation}
 dS_{\mathcal Q}^{\rm TR} = \frac{dU}{T_{\mathcal Q}}.
\end{equation}
In order to maintain a strictly constant heat capacity, the $S_{\mathcal Q}$ 
of a TR must be linear in $U$. 

\section{Concluding remarks}

We have reviewed the two model-free derivations of the Boltzmann
factor: constant-$T$ and small-$E_s$ derivations. As an extension of
the constant-$T$ derivation, it is proposed that the
constant heat capacity derivation, in which the heat capacity
$k/{\mathcal Q}$ of the environment is assumed to be exactly constant
and given by the real parameter of ${\mathcal Q}$. It is shown
that for any system in thermal contact with such a constant heat capacity
environment, the equilibrium distribution is identical to a
power-law Tsallis distribution. 
This fact is interesting because Tsallis distribution is obtained
without resort to Tsallis' entropy of Eq. \eqref{Sq}, which is
an entropy {\it \'a la} Gibbs.
In addition the Clausius' definition of thermodynamic entropy is
consistent with the standard Boltzmann entropy of a constant heat capacity
environment, whose temperature linearly depends on internal energy. 

Finally, let me comment on a connection of the constant heat capacity
with (multi-)fractal energy spectra. For a (multi-)fractal energy
spectrum, it is shown \cite{vall98,carp00} that its integrated density
of state is well fitted to a power-law as $\Omega(E) \propto E^{d_{\rm
E}}$, where $d_{\rm E}$ is the fitting exponent of the power-law fit.
Consequently the average heat capacity is constant, $\langle C \rangle
=k d_{\rm E}$, hence ${\mathcal Q}=1/d_{\rm E}$. The key point is that
the ${\mathcal Q}$ deviates from $0$ since $d_{\rm E}$ is not large.
On the contrary, for a classical gas environment \cite{pren99} 
$\Omega_{\rm gas}(E) \propto E^f$ with $f$ is the degree of freedom of 
the gas environment,
the corresponding ${\mathcal Q}_{\rm gas}=1/f$ tends to $0$ 
when $f$ becomes infinite.
Consequently any system thermally contact with the gas environment obeys
a Boltzmann distribution.
Thus an environment which has a (multi-)fractal energy
spectrum may provide a constant heat capacity environment with 
${\mathcal Q} \ne 0$.
In some case, but not always, $d_{\rm E}$ is related
to the properties of the (multi-)fractal energy spectra. For example, for
the energy spectrum of the Cantor set, $d_{\rm E}$ equals to the
fractal dimension ($=\ln 2/\ln 3)$ of the Cantor set.

%

\end{document}